# The Impact of COVID-19 Vaccination Delay: A Modelling Study for Chicago and NYC Data

Vinicius V. L. Albani[1], Jennifer Loria[2,3], Eduardo Massad[4,5], and Jorge P. Zubelli[6]

February 5, 2020

**Abstract**

**Background:** By the beginning of December 2020, some vaccines against COVID-19 already presented efficacy and security, which qualify them to be used in mass vaccination campaigns. Thus, setting up strategies of vaccination became crucial to control the COVID-19 pandemic.

**Methods:** We use daily COVID-19 reports from Chicago and NYC from 01-Mar2020 to 28-Nov-2020 to estimate the parameters of an SEIR-like epidemiological model that accounts for different severity levels. To achieve data adherent predictions, we let the model parameters to be time-dependent. The model is used to forecast different vaccination scenarios, where the campaign starts at different dates, from 01-Oct-2020 to 01-Apr-2021.

---

[1] Federal University of Santa Catarina, 88.040-900 Florianopolis, Brazil, v.albani@ufsc.br

[2] Instituto de Matemática Pura e Aplicada, Rio de Janeiro, Brazil, jennyls@impa.br

[3] Universidad de Costa Rica, San José, Costa Rica

[4] School of Medicine, University of São Paulo and LIM01-HCFMUSP, São Paulo, Brazil

[5] School of Applied Mathematics, Fundação Getúlio Vargas, Rio de Janeiro, Brazil, eduardo.massad@fgv.br

[6] Mathematics Department, Khalifa University, Abu Dhabi, UAE, jorge.zubelli@ku.ac.ae

To generate realistic scenarios, disease control strategies are implemented whenever the number of predicted daily hospitalizations reaches a preset threshold.

**Results:** The model reproduces the empirical data with remarkable accuracy. Delaying the vaccination severely affects the mortality, hospitalization, and recovery projections. In Chicago, the disease spread was under control, reducing the mortality increment as the start of the vaccination was postponed. In NYC, the number of cases was increasing, thus, the estimated model predicted a much larger impact, despite the implementation of contention measures.

The earlier the vaccination campaign begins, the larger is its potential impact in reducing the COVID-19 cases, as well as in the hospitalizations and deaths. Moreover, the rate at which cases, hospitalizations and deaths increase with the delay in the vaccination beginning strongly depends on the shape of the incidence of infection in each city.

# 1  Introduction

Previous pandemics have demonstrated that, as a general rule, pharmaceutical interventions are less important than non-pharmaceutical intervention in controlling the infection, however, there is a possibility that this will not be the case with the vaccines against COVID-19 [1–3].

Some few months after the emergence of SARS-CoV-2 in China, several academic laboratories and pharmaceutical industries around the world started the development of more than 100 types of different vaccines, short-circuiting in less than one year the usual time frame of new vaccines development and testing of around ten years [3,4].

There is, therefore, an enormous variety of COVID-19 vaccines being developed. As of November 2020, there were 48 vaccines in clinical trials and 146 candidate vaccines in pre-clinical evaluation [5]. Of these, 12 vaccines were in the pipeline, of which ten were in Phase 3 of clinical trials (four have already completed this phase) and two were in Phase 2 [5]. In the US, three vaccines completed Phase 3 trials, namely, Moderna, Pfizer, and AstraZeneca, and two were still in Phase 3 [5].

In order to have a significant impact on the course of the pandemic, however, safe and effective vaccines have to emerge in less time it would take the affected populations to reach natural herd immunity [3]. Therefore, an unprecedented time-schedule to roll out any effective vaccine is urgently needed.

In December 2020, the CDC proposed the Phase 1 allocation schedule of vaccination, covering an estimated 264 million people in about 25 weeks from the beginning of vaccination. Phase 1a would cover 21 million health personnel and three million nursing residents. Phase 1b would cover 87 million essential workers, 100 million persons with risky medical conditions and 53 million adults older than 65 years of age [6]. This ambitious rolling out plan, however, is way behind schedule. By 7-Jan-2021, only about five million people have been vaccinated [7].

We quantify the delay impact in vaccination deployment under different scenarios using publicly available data. This is done by implementing a novel Susceptible-Exposed-Infective-Recovered-like (SEIR) model that accounts for the different levels of disease severity, asymptomatic infection, age range, and regime changes in disease spread. The model captures well the time evolution of the outbreak leading to the forecast of realistic scenarios. It is tested with publicly available data from Chicago and NYC confirming adherence to historical data. We observe that according to the disease-spread control level, the impact of postponing a mass vaccination campaign is considerable. Reopening strategies after lockdown are also accounted for in our study.

It is worth mentioning that the politicization of the vaccination in many countries, the polemic around safety and efficacy of the candidate vaccines and the anti-vaccination groups campaigning against the vaccine are all contributing to hesitancy [8] and an inevitable delay in vaccination in many places around the world (not to mention the technical hurdles to roll out billions of doses necessary to control the SARS-CoV-2 pandemic). All these issues make the estimation of the death toll caused by vaccination delay of utmost importance.

## 2 Materials and Methods

This section presents the epidemiological model as well as the estimation techniques used to calibrate the model parameters from observed cases of COVID-19.

## 2.1  The Epidemiological SEIR-like Model

The proposed epidemiological model accounts for the distribution of the population into $n$ age ranges. For the $i$th age range, the corresponding group of individuals is further classified into nine compartments, namely, susceptible (S), vaccinated (V), exposed (E), asymptomatic and infective ($I_A$), mildly infective ($I_M$), severely infective ($I_S$), critically infective ($I_C$), recovered ($R$), and deceased ($D$). All infective individuals that are symptomatic but do not need to be hospitalized are considered mildly infective. By severely infective we mean those individuals that were admitted to a regular hospital bed. Those individuals that were admitted to an intensive care unit (ICU) are considered as critically infective. Only susceptible individuals are considered vaccinated, which means that if someone is vaccinated after being exposed, then he or she will pass to the asymptomatic or mildly infective compartments. Before presenting the model, let us introduce the vector notation, i.e.,

$$\mathbf{S} = [S^1, \cdots, S^n]^T,$$

where $S^i$ ($i = 1, \cdots, n$) represents the susceptible individuals in the $i$th age range. $\mathbf{E}, \mathbf{V}, \mathbf{I}_A, \mathbf{I}_M, \mathbf{I}_S, \mathbf{I}_C, \mathbf{R},$ and $\mathbf{D}$ are defined similarly. Also consider the tensor product between two $n$-dimensional vectors, defined as

$$\mathbf{X} \colon \mathbf{Y} = [x^1 y^1, \cdots, x^n y^n]^T.$$

Thus, the movement between the model compartments is determined by the following system of ordinary differential equations:

$$\dot{\mathbf{S}} = -\mathbf{S} \colon (\beta_A \mathbf{I}_A + \beta_M \mathbf{I}_M + \beta_S \mathbf{I}_S + \beta_C \mathbf{I}_C) - \nu \colon \mathbf{S}, \quad (1)$$
$$\dot{\mathbf{V}} = \nu \colon \mathbf{S}, \quad (2)$$
$$\dot{\mathbf{E}} = \mathbf{S} \colon (\beta_A \mathbf{I}_A + \beta_M \mathbf{I}_M + \beta_S \mathbf{I}_S + \beta_C \mathbf{I}_C) - \sigma \mathbf{E}, \quad (3)$$
$$\dot{\mathbf{I}}_A = (1-p)\sigma \mathbf{E} - \gamma_{R,A} \colon \mathbf{I}_A, \quad (4)$$

$$\dot{I}_M = p\sigma E - (\gamma_{R,M} + \alpha_S) \cdot I_M, \quad (5)$$
$$\dot{I}_S = \alpha_S \cdot I_M - (\gamma_{R,S} + \alpha_C) \cdot I_S, \quad (6)$$
$$\dot{I}_C = \alpha_C \cdot I_S - (\gamma_{R,C} + \delta_D) \cdot I_C, \quad (7)$$
$$\dot{R} = \gamma_{R,A} \cdot I_A + \gamma_{R,M} \cdot I_M + \gamma_{R,S} \cdot I_S + \gamma_{R,C} \cdot I_C, \quad (8)$$
$$\dot{D} = \delta_D \cdot I_C. \quad (9)$$

The schematic representation of the model can be found in Figure 1.

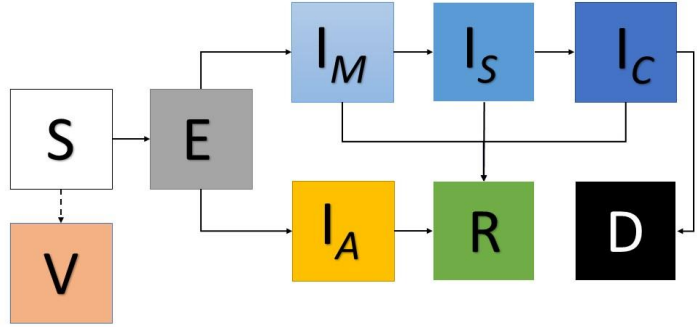

Figure 1: Schematic representation of the epidemiological model of Eqs. (1)-(9).

The time-dependent transmission parameters for asymptomatic, mildly, severely, and critically infective individuals are denoted, respectively, by $\beta_A$, $\beta_M$, $\beta_S$, and $\beta_C$. The rate of vaccination is $\nu$, which is given by the product of the daily rate of vaccination of susceptible individuals by the effectiveness of the used vaccine. The meantime from contagion to become infective is the inverse of the parameter $\sigma$. The recovery rate of mildly, severely, critical and asymptomatic infective individuals are denoted by $\gamma_{R,M}$, $\gamma_{R,S}$, $\gamma_{R,C}$, and $\gamma_{R,A}$, respectively. The rates of hospitalization and ICU admission are denoted by $\alpha_S$ and $\alpha_C$, respectively. According to the World Health Organization, only people in severe conditions generally die by COVID-19, thus, the corresponding death rate is $\delta_D$ [9].

The unknown parameters are $\beta_A$, $\beta_M$, $\beta_S$, and $\beta_C$, as well as the initial number of mildly and asymptomatic infective individuals, that shall be estimated from the daily numbers of infections. In order to reduce the number of unknowns, we assume that

$$\beta_S = a\beta_M, \beta_C = b\beta_M, \text{ and } \beta_A = c\beta_M, \qquad (10)$$

with $a = 0.1$, $b = 0.01$, and $c = 0.58$, which means that the infection rate of hospitalized, in ICU and asymptomatic individuals are 10%, 1% and 58%, respectively, of the transmission rate of those ones in the mildly infective compartment [10,11]. The mean time between infection and becoming infective is set to 5.1 [12]. The proportion of exposed individuals becoming mildly infective is $p$, which is set to 0.83 [11]. The recovery rates of mildly, severely, and critically infective individuals are simply set as one minus the rates of hospitalization, ICU admission, and death, respectively. All the asymptomatic individuals will recover in 14 days, so, $\gamma_{R,A} = 14^{-1}$. The rate of ICU admission is set as $\alpha_C = 0.4$ [13]. The hospitalization and death rates are time-dependent and defined as follows:

$$\alpha_S(t) = \frac{H(t)}{I(t-1)} \quad \text{and} \quad \delta_D(t) = \frac{D(t)}{\alpha_C H(t-1)} \qquad (11)$$

where $I, H,$ and $D$ represent the time series of seven-day moving averages of daily numbers of infections, hospitalizations and deaths, respectively.

If the number of age ranges $n$ is larger than one, the entries of the matrix $\beta_M$ are defined as:

$$[\beta_M]_{ii} = \beta^i(t)b_i, \ [\beta_M]_{ij} = \frac{p_j}{2}\big(\beta^i(t)b_i + \beta^j(t)b_j\big), \ i \neq j \ (i,j = 1,\cdots,n),$$

where $\beta^i(t)$ ($i = 1,\ldots,n$) are time-dependent scalar coefficients, and $b_i$, as well as $p_i$ ($i = 1,\ldots,n$) are time-independent [10]. $\beta^i(t)$ represents the time-dependent part of the transmission coefficient for the $i$th age range, whereas $b_i$ is the time-independent part. They, respectively, capture the short-term and the long-term pattern of the disease spread. The parameter $p_i$ represents the probability of any individual from the $j$th age range to interact with an individual from the $(i + j - 1)$th age range. In this case, we set $p_1 = 1$. Under these assumptions, the number of unknown parameters in $\beta_M$, for each $t$, is $2n$, instead of $n^2 - n$.

## 2.2 Estimation Techniques

As aforementioned, the set of unknown parameters is composed by the initial number of individuals at the mildly infective individuals $\mathbf{I}_M(0) = [I_M^1(0), \cdots, I_M^n(0)]^T$, the time-dependent components of the matrix $\beta_M$, $\vec{\beta}(t) = [\beta_1(t), \cdots, \beta_n(t)]^T$, and the time-independent components of $\beta_M$, $\mathbf{b} = [b_1, \cdots, b_n]^T$ and $\mathbf{p} = [1, p_2, \cdots, p_n]^T$. Let $\mathfrak{I}^i = \{\mathcal{I}^i(t_1), \ldots, \mathcal{I}^i(t_N)\}$ represent the time series of seven-day moving averaged version of daily reports of COVID-19 infections for the $i$th age range. The estimation procedure is performed in two steps. In the first one, the time-independent parameters, $\mathbf{I}_M(0)$ and $\mathbf{b}$, are estimated and, in the second one, $\vec{\beta}(t)$ is calibrated.

The estimated $\mathbf{I}_M(0)$ and $\mathbf{b}$ are minimizers of the objective function below, which is closely related to the so-called posterior density of the model parameters, given the set of reports [14].

$$\Pi(\mathbf{I}_M(0), \mathbf{b}, \mathbf{p} | \mathfrak{I}^1, \cdots, \mathfrak{I}^n) = L(\mathfrak{I}^1, \cdots, \mathfrak{I}^n | \mathbf{I}_M(0), \mathbf{b}, \mathbf{p}) + \alpha Pr(\mathbf{I}_M(0), \mathbf{b}, \mathbf{p}), (12)$$

where $\alpha > 0$ is a penalty parameter,

$$L(\mathfrak{I}^1, \cdots, \mathfrak{I}^n | \mathbf{I}_M(0), \mathbf{b}, \mathbf{p}) = \sum_{i=1}^{n} \sum_{l=1}^{N} [\mathcal{I}^i(t_l) \log(\sigma E^i(t_l)) - \sigma E^i(t_l) - \log(\mathcal{I}^i(t_l)!)]$$

is the data misfit function with $\log(\mathcal{I}^i(t_l)!)$ approximated by the Stirling's formula

$$\log(\mathcal{I}^i(t_l)!) \approx \frac{1}{2} \log(2\pi \mathcal{I}^i(t_l)) + \mathcal{I}^i(t_l) \log(\mathcal{I}^i(t_l)) - \mathcal{I}^i(t_l),$$

and

$$Pr(\mathbf{I}_M(0), \mathbf{b}, \mathbf{p}) = \|(\mathbf{I}_M(0), \mathbf{b}, \mathbf{p}) - (\mathbf{I}_M^0(0), \mathbf{b}^0, \mathbf{p}^0)\|_{\ell_2}^2$$

is the penalty term, with $(\mathbf{I}_M^0(0), \mathbf{b}^0, \mathbf{p}^0)$ a set of *a priori* parameters.

The set of time-dependent components of $\beta_M$, $\vec{\beta}(t)$ is estimated by minimizing the function below, for each $t_l$ ($l = 1, \cdots, N$):

$$\Pi(\vec{\beta}(t_l)|\mathcal{I}^1(t_l),\ldots,\mathcal{I}^n(t_l), \mathbf{I}_M(0), \mathbf{b}, \mathbf{p})$$

$$= \sum_{i=1}^{n} [\mathcal{I}^i(t_l) log(\sigma E^i(t_l)) - \sigma E^i(t_l) - log(\mathcal{I}^i(t_l)!)]$$

$$+ \alpha \|\vec{\beta}(t_l) - \vec{\beta}(t_{l-1})\|_{\ell_2}^2.$$

The minimization of the objective functions above is performed by a gradient-based technique [15]. Confidence intervals (CIs) are generated by bootstrapping, based on a set of 200 samples [16].

## 3    Results

This section presents the comparison between model predictions and reported data after calibration, as well as vaccination scenarios created with the calibrated model using data from Chicago and New York City (NYC).

## 3.1  Estimation Results

The parameters of the epidemiological model of Eqs. (1)-(9) are estimated from seven-day moving average time-series of daily new infections from Chicago and NYC. The time-series of daily reports of COVID-19 infections, as well as related hospitalizations and deaths for Chicago and NYC, are available from public resources [17,18]. Recent census data containing the total population of the considered cities and their distributions by age ranges were also used [19].

During the model estimation, the time series were divided into sets of consecutive 20 days. Besides the set corresponding to the beginning of the COVID-19 outbreak in these cities, for each 20-day dataset, **b** and $\vec{\beta}(t)$ are estimated. We start by not distributing the population into age ranges, which means that *n* is set to 1 in the model of Eqs. (1)-(9). The time-dependent effective reproduction number denoted by $\mathcal{R}(t)$ is evaluated through the next-generation matrix technique [20,21].

Model prediction using estimated parameters of daily new cases, hospitalizations and deaths, as well as the corresponding reported numbers for Chicago and NYC for the period 01-Mar-2020 to 28-Nov-2020 can be found in Figures 2-3, respectively. The corresponding effective reproduction numbers are also presented.

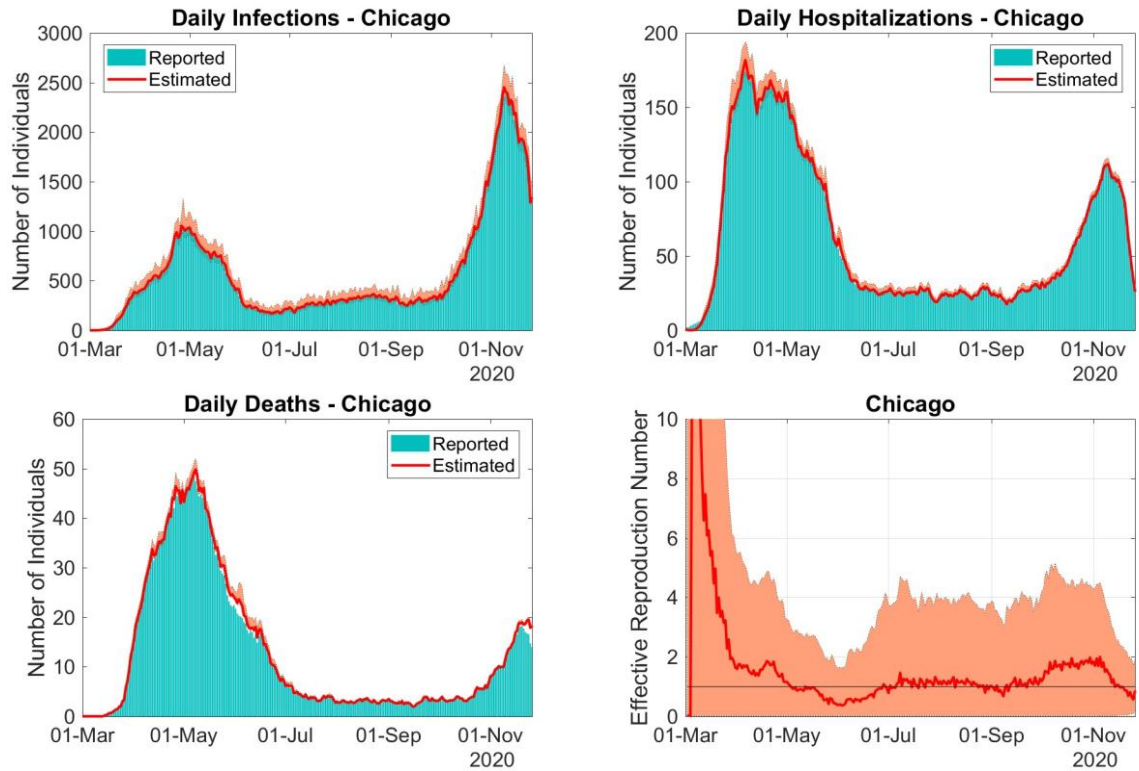

Figure 2: Model predictions of infections, hospitalizations and deaths (solid lines), using data from Chicago. The bars represent the reports and the envelopes are 90% CIs. The corresponding seven-day moving average of the time-dependent basic reproduction rate is depicted in the bottom right panel.

For both cities, the model predictions of daily new cases, hospitalizations and deaths (in Figure 2, top left and right, as well as bottom left panels, respectively) are adherent to the reports. It is explained by the effectiveness of the calibration procedure, and the use of the hospitalization and death rates defined in Eq. (11). We decided to present the seven-day moving average of $\mathcal{R}(t)$ since it is less fuzzy, allowing to see the qualitative trend of the spread dynamics, such as the effectiveness of control measures. The periods when control measures effectively reduced the number of new COVID-19 infections are illustrated by the graph of $\mathcal{R}(t)$, where, during such dates, its value remained below one (solid horizontal line).

## 3.2  Vaccination Scenarios

Let us consider that a vaccination campaign is implemented in Chicago and NYC. The vaccine is 95% effective. Firstly, during the campaign, on each day, 1% of the susceptible population is immunized, until the number of susceptible individuals is less than 40%. This threshold was chosen based on an estimate of the proportion of US citizens that accept to get a vaccine against COVID-19 [22].

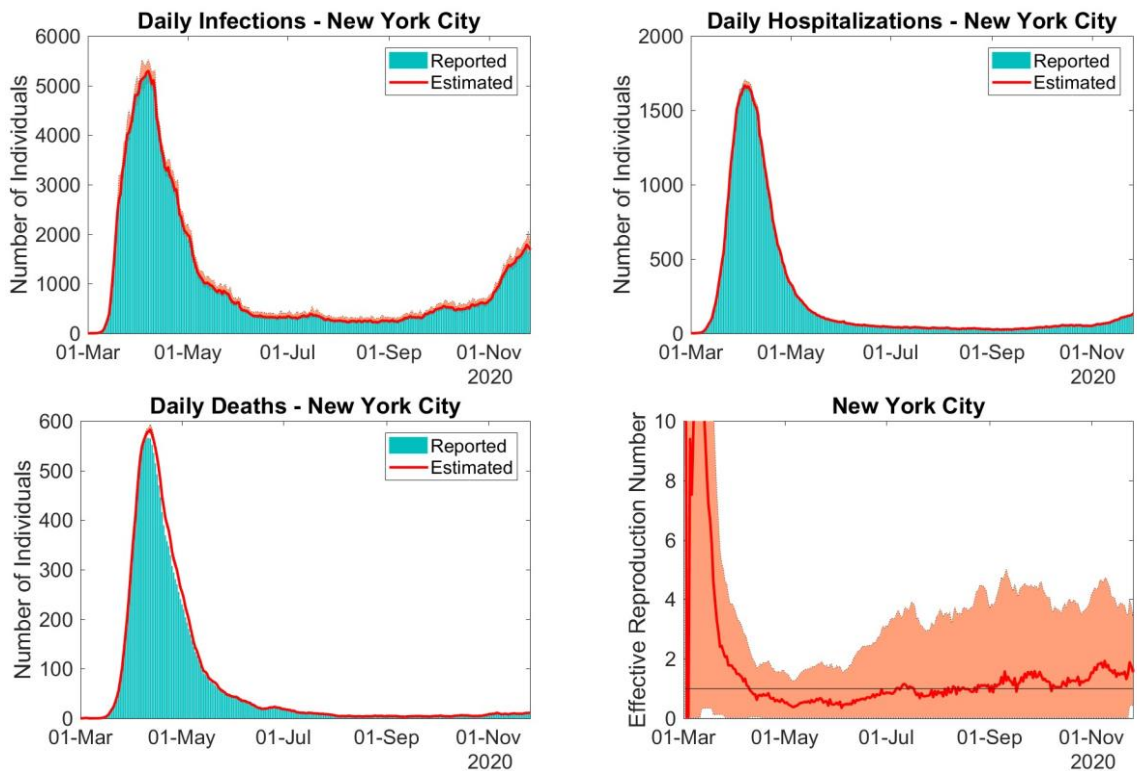

Figure 3: Model predictions of infections, hospitalizations and deaths (solid lines), using data from NYC. The bars represent the reports and the envelopes are 90% CIs. The corresponding seven-day moving average of the time-dependent basic reproduction rate is depicted in the bottom right panel.

In order to forecast scenarios, the time-dependent parameters are extended to the forecast period by repeating the average of the values estimated in the last ten days of the calibration period. To avoid unrealistic numbers, whenever the predicted number of daily hospitalizations reaches the value 300, the time-dependent transmission coefficient $\beta(t)$ is set to the average of the values estimated in the period 07-Sept-2020 to 16-Sept-2020, when the disease spread was controlled. In this period, the effective reproduction numbers in Figures 2-3 were close to the value one, indicating that the disease spread was under control.

The vaccination campaign is set in the period 01-Oct-2020 to 31-May-2021, starting on different dates, but finishing at 31-May-2020. Table 1 shows the accumulated numbers of infections, hospitalizations and deaths corresponding to the different starting dates.

The evolution of the number of accumulated deaths with respect to the starting date of the vaccination campaign can be found in Figure 4. The increasing number of deaths, as the beginning of the campaign is postponed, also illustrate that vaccination must start as soon as possible.

| Starting Date | Cases | Hospitalizations | Deaths | Total Vaccinated |
|---|---|---|---|---|
| 01-Oct-2020 | 55,628 (50,666–62,342) | 2,345 (2,255–2,462) | 317 (305–331) | 1,474,347 (1,463,025–1,475,247) |
| 01-Nov-2020 | 115,236 (102,557–131,706) | 4,177 (3,993–4,411) | 683 (648–726) | 1,408,332 (1,405,379–1,410,475) |
| 01-Dec-2020 | 144,489 (123,310–172,791) | 4,732 (4,466–5,075) | 881 (814–966) | 1,378,253 (1,359,768–1,394,343) |
| 01-Jan-2021 | 159,687 (131,278–200,399) | 4,976 (4,639–5,428) | 983 (886–1,113) | 1,367,845 (1,332,090–1,388,244) |
| 01-Feb-2021 | 167,904 (134,615–219,067) | 5,108 (4,721–5,648) | 1,037 (920–1,204) | 1,351,928 (1,309,981–1,385,688) |
| 01-Mar-21 | 171,976 (135,916–230,405) | 5,173 (4,757–5,772) | 1,064 (935–1,255) | 1,349,108 (1,302,692–1,374,433) |
| 01-Apr-2021 | 174,340 (136,523–238,305) | 5,211 (4,775-5,853) | 1,080 (942–1,289) | 1,030,436 (1,008,001–1,042,678) |

Table 1: Accumulated numbers of infections, hospitalizations and deaths in Chicago, when the vaccination campaign starts at different dates. Such values are based on model predictions using the estimated parameters.

| Starting Date | Cases | Hospitalizations | Deaths | Total Vaccinated |
|---|---|---|---|---|
| 01-Oct-2020 | 49,336 (40,942–59,625) | 3,314 (3,036–3,634) | 360 (335–387) | 4,669,601 (4,668,032–4,671,154) |
| 01-Nov-20 | 133,301 (97,217–178,086) | 8,567 (7,324–10,000) | 876 (761–1,007) | 4,570,986 (4,563,555–4,610,135) |
| 01-Dec-2020 | 310,265 (177,628–453,783) | 19,810 (14,957–25,634) | 1,933 (1,480–2,475) | 4,432,224 (4,333,769–4,535,830) |
| 01-Jan-21 | 543,672 (282,852–1,030,119) | 34,682 (25,689–47,443) | 3,329 (2,488–4,522) | 4,190,610 (3,795,860–4,417,414) |
| 01-Feb-2021 | 908,557 (407,618–2,006,687) | 57,915 (36,641–88,064) | 5,507 (3,515–8,329) | 3,823,141 (2,935,676–4,284,318) |
| 01-Mar-21 | 1,303,397 (504,256–3,029,771) | 83,026 (46,905–133,982) | 7,855 (4,473–12,618) | 3,431,009 (2,030,837–4,145,321) |
| 01-Apr-2021 | 1,731,320 (584,568–3,977,235) | 110,098 (57,672–179,733) | 10,358 (5,470–16,860) | 2,743,885 (1,211,852–3,146,459) |

Table 2: Accumulated numbers of infections, hospitalizations and deaths in NYC, when the vaccination campaign starts at different dates. Such values are based on model predictions using the estimated parameters.

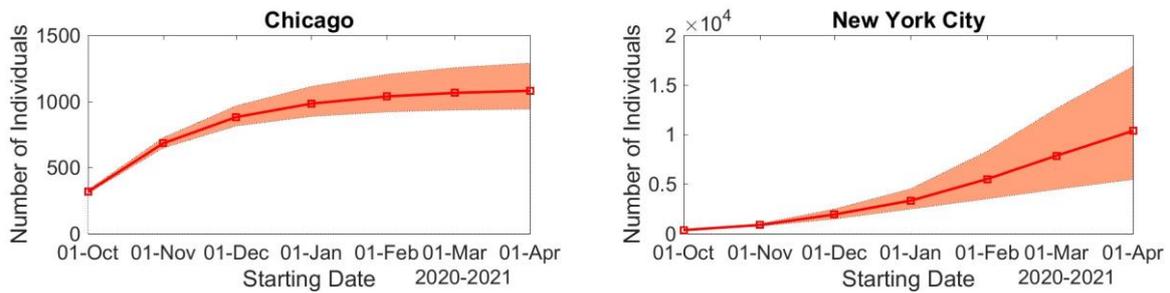

Figure 4: Evolution of the model predicted accumulated deaths in Chicago (left) and in NYC (right) with respect to the starting date of the vaccination campaign. The envelopes are 90% CIs.

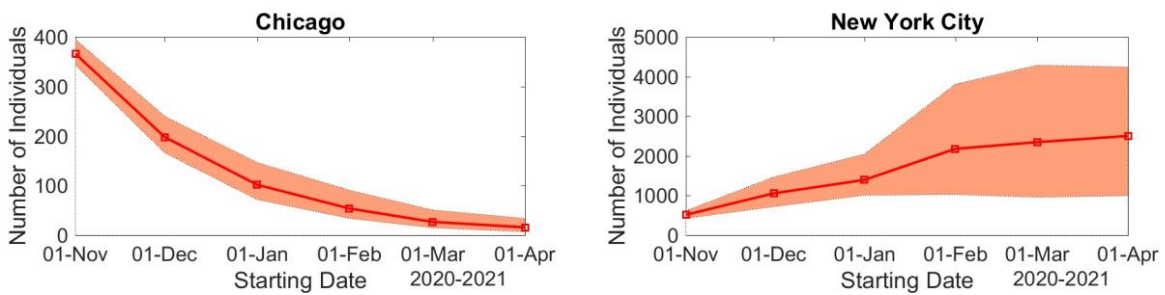

Figure 5: Model predictions for the increment in the accumulated deaths by postponing the starting date of the vaccination campaign in Chicago (left) and in NYC (right). The envelopes are 90% CIs.

An example using a vaccination strategy that accounts for age range can be found in the supplement. The corresponding conclusions and results are similar to the ones above.

## 4    Discussion

In this paper, to generate vaccination scenarios, we propose an SEIR-like model that accounts for the different levels of disease severity, asymptomatic infection, age range, and regime changes in disease spread as times goes by. The model parameters are calibrated from reports of daily COVID-19 infections, as well as published reports. We end-up with a modeling tool that captures well the time evolution of the outbreak, reproducing the empirical data with remarkable accuracy, helping to forecast realistic scenarios. Such features are illustrated using publicly available data from Chicago and NYC.

Depending on, whether the disease spread is under control or not, that is, whether the daily incidence curve of infection is increasing or decreasing, the impact of postponing the beginning of a mass vaccination campaign is considerable. As expected, such impact is more serious in regions where the incidence curve is increasing than in cities where the infection is controlled.

We use different strategies and consider the implementation of contention measures, as the daily reports of hospitalizations reach a threshold. Reopening strategies after lockdown are also accounted for in our study.

The model has some important limitation worth mentioning. First, it assumes that 60% of susceptible are vaccinated with a 95% efficacy vaccine in a short period of time at a rate of 1% per day. Although this scenario is logistically feasible, it is a daunting task.

The current scenario of the pandemic, in which new variants of SARS-CoV-2 are emerging in some countries, should be considered in the simulation of future vaccination models [23]. However, there is not enough empirical evidence of the repercussion of these new variants of the vaccine efficacy.

Finally, we should point out that the SARS-CoV-2, like any other viruses, is evolving, with new strains showing increased transmissibility. It should be expected, however, that its case fatality rate (or virulence) should be decreasing with time. This is a general rule in the evolution of pathogens which helps them to increase their basic reproduction number [24]. If this is the case, then it is possible to predict that in few years, COVID-19 tends to be a mild disease as other coronaviruses, like OC-43 which probably caused the so-called "Russian flu" in 1889 and nowadays is responsible for about 10% of the common cold [25]. The future of vaccines against SARS-CoV-2, therefore, will very much depend on the virulence the virus will eventually evolve.

## 5 ICMJE Statement

All authors attest they meet the ICMJE criteria for authorship.

**Contributors**

VA, EM and JZ proposed the mathematical model. VA and JL performed numerical simulations. VA and JL analyzed the data. All the authors contributed to the writing of the article. EM and JZ critically revised the manuscript. All the authors had full access to all data used in the study and take responsibility for the accuracy of the data analysis. All authors revised and approved the final version of the article.

**Declaration of interests**

All authors declare no competing interests.

**Data sharing**

The data that support the findings of this study are available from publicly available sources [17,18]. The numerical scripts used to generate the simulated scenarios can be found in the GitHub repository https://github.com/JennySorio/Vaccination_Scenarios.

**Funding**

This work was supported by the Conselho Nacional de Desenvolvimento Científico e Tecnológico (CNPq) [grant numbers 305544/2011-0 and 307873/2013-7], the Fundação

Butantan [grant number 01/2020], the Fundação Carlos Chagas Filho de Amparo à Pesquisa do Estado do Rio de Janeiro [grant number E-26/202.927/2017], and the Universidad de Costa Rica (UCR) [grant number OAICE-CAB-02-022-2016].

# Appendix A

## A.1 One Additional Example

We now simulate vaccination campaigns, starting on different dates, where people over 80 years old are immunized one month earlier than individuals in other age ranges. In addition, people under 18 years old are not vaccinated. Again, we assume 95% of effectiveness of the vaccine and the rate of vaccination of the population in the $i$th age range is 1%.

| Starting Date | Cases | Hospitalizations | Deaths | Total Vaccinated |
|---|---|---|---|---|
| 01-Nov-20 | 127,262 (104,535–159,129) | 7,642 (6,800–8,671) | 650 (578–731) | 4,282,170 (4,200,732–4,377,917) |
| 01-Dec-2020 | 236,855 (172,234–328,778) | 14,185 (11,614–17,497) | 1,194 (980–1,460) | 4,168,087 (4,081,642–4,259,118) |
| 01-Jan-21 | 404,531 (251,913–633,723) | 24,304 (17,882–32,976) | 2,033 (1,506–2,732) | 4,005,264 (3,787,083–4,176,623) |
| 01-Feb-2021 | 632,154 (338,744–1,084,018) | 38,133 (25,329–55,809) | 3,188 (2,134–4,626) | 3,663,073 (3,343,475–3,831,779) |
| 01-Mar-21 | 876,347 (419,642–1,559,624) | 53,220 (32,802–80,925) | 4,472 (2,777–6,759) | 3,014,098 (2,705,315–3,232,245) |
| 01-Apr-2021 | 1,128,861 (497,551–2,005,552) | 68,724 (40,222–105,271) | 5,769 (3,403–8,813) | 2,175,741 (1,861,952–2,402,590) |

Table 3: Accumulated numbers of infections, hospitalizations and deaths in NYC, when the vaccination campaign starts at different dates. Such values are based on model predictions using the estimated parameters obtained in Section 3.1.

Table 3 presents the accumulated numbers of COVID-19 cases, hospitalizations and deaths, as well as of immunized individuals during the period 01-Nov-2020 to 31-May-2021 for NYC. As the starting date of the campaign is delayed, there is a remarkable increase in the accumulated numbers.

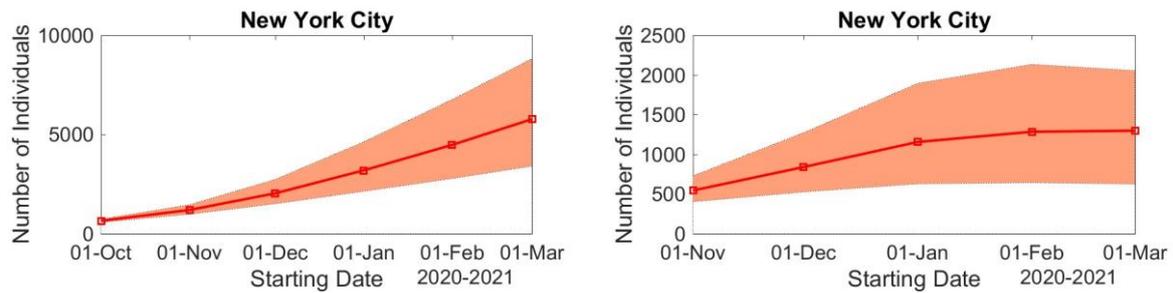

Figure 6: Right: Evolution of the model predicted accumulated deaths in NYC with respect to the starting date of the vaccination campaign. Left: increment in the accumulated deaths by postponing the starting date of the vaccination campaign. The envelopes are 90% CIs.

The left panel in Figure 6 shows the evolution of the accumulated number of deaths as a function of the vaccination campaign starting date. The right panel shows the increment in the number of deaths for each starting date, in comparison to starting vaccination one month earlier.